\documentclass[conference]{IEEEtran}
\usepackage{graphicx}
\usepackage{hyperref}

\usepackage{cite}

\ifCLASSINFOpdf
\else
\fi
\begin{document}
%
\title{
%
Reliability of Dynamic Load Scheduling with Solar Forecast Scenarios}
%




%

\author{\IEEEauthorblockN{Abdulelah H Habib\IEEEauthorrefmark{1},
Zachary K Pecenak\IEEEauthorrefmark{1},
Vahid R Disfani\IEEEauthorrefmark{1}, 
Jan Kleissl\IEEEauthorrefmark{1} and
Raymond A. de Callafon\IEEEauthorrefmark{1}}
\IEEEauthorblockA{\IEEEauthorrefmark{1} Jacobs School of Engineering, Mechanical and Aerospace Engineering\\
University of California, San Diego, La Jolla, CA 92093\\ Email: ahhabib, zpecenak, disfani, jkleissl, callafon@ucsd.edu}
}


\maketitle

\begin{abstract}
This paper presents and evaluates the performance of an optimal scheduling algorithm that selects the on/off combinations and timing of a finite set of dynamic electric loads on the basis of short term predictions of the power delivery from a photovoltaic source. In the algorithm for optimal scheduling, each load is modeled with a dynamic power profile that may be different for on and off switching. Optimal scheduling is achieved by the evaluation of a user-specified criterion function with possible power constraints. The scheduling algorithm exploits the use of a moving finite time horizon and the resulting finite number of scheduling combinations to achieve real-time computation of the optimal timing and switching of loads. The moving time horizon in the proposed optimal scheduling algorithm provides an opportunity to use short term (time moving) predictions of solar power based on advection of clouds detected in sky images. Advection, persistence, and perfect forecast scenarios are used as input to the load scheduling algorithm to elucidate the effect of forecast errors on mis-scheduling. The advection forecast creates less events where the load demand is greater than the available solar energy, as compared to persistence. Increasing the decision horizon leads to increasing error and decreased efficiency of the system, measured as the amount of power consumed by the aggregate loads normalized by total solar power. For a standalone system with a real forecast, energy reserves are necessary to provide the excess energy required by mis-scheduled loads. A method for battery sizing is proposed for future work.

\end{abstract}


%
\IEEEpeerreviewmaketitle

\section{Introduction}

Power variability is one of the main obstacles facing renewable energy expansion. Solar and wind are of main concern since their electrical production is proportional to the variability associated with wind and solar resources\cite{nrelvar}. This poses an issue for current electrical infrastructure which was designed around variability in demand, not generation. Transient local demand changes over seconds or minutes are for the most part small and spatially uncorrelated resulting a relatively steady demand profile. Over several hours, loads can change substantially, but these changes in load have a tendency to be more predictable. This is manifested through daily patterns of morning load pickup and evening load drop-off highly correlated with human activity\cite{c22,c21}. 

Wind and solar generation, on the other hand, is variable. An individual wind turbine or solar plant can ramp from full to less than half of production in a minute. On the other hand the aggregate variability of multiple turbines at the same site or even all renewable generators in a balancing area is relatively much less \cite{nrelvar}. Still such variability is not entirely predictable and therefore causes uncertainty in projecting power output minutes to days ahead. 

Variability and uncertainty are more critical in standalone or island mode applications where a high penetration of renewable power sources ramping near synchronously may create power variability that is large enough to cause substantial power quality and/or grid economics issues. An approach to solve this problem is  to increase scheduling and adjustments to controllable loads to ``load follow''  wind and solar generation on the grid \cite{market}. Examples of controllable loads are devices such as air conditioners and refrigerators with a temperature dead band that effectively creates a thermal storage reservoir. The load power of these type of machines can be changed temporarily while respecting the demands of the end user \cite{c20}. Another example is scheduling of more intermittent loads such as water pumps, which can be adjusted to accommodate power variability, claim optimal power usage, and decrease power losses. The control and scheduling of such loads benefits from supply and load forecast \cite{UncertaintyForecast}.\\
\indent Load scheduling has been applied in many fields, such as thermal loads, residential appliances, and EV charging \cite{c8}. In \cite{c3}, a case study was implemented to accomodate wind power variability through EV charging. Another example for household appliances scheduling was demonstrated in \cite{c2, Householdloadscheduling}. From the supply side, \cite{c9} discussed new dispatch methodology to power and control a hybrid wind turbine and battery system.
In \cite{c2}, more work was done toward game theory and customers' effect on the grid. \\ 
\indent Model Predictive Control (MPC) was used in most approaches\cite{c12,c13} to compute optimal control or scheduling signals for the load. Typically, in MPC a constraint (quadratic) optimization problem is solved iteratively over a finite $N$ and moving time $t$ horizon from $t=k$ till $t=k+N-1$ to compute an optimal control signal in real time, denoted here by $w(k)$ at time $t=k$. Countless examples of innovative MPC based approaches for load scheduling, grid tied storage systems or to maintain Voltage stability can be found in e.g. \cite{c3}, \cite{c14}, \cite{c15} or \cite{c18}. Although MPC approaches are extremely powerful in computing optimal control signals over a moving but finite time horizon, typically the control signal $w(k)$ is allowed to attain any real value during the optimization, see e.g. \cite{c16,c17,c19}. Unfortunately, a real-valued control signal $w(k)$ would require distinctive loads on an electric grid to operate at fractional load demands. Although this can be implemented by electric storage systems or partial or pulse width modulation of loads \cite{c15}, (non-linear) dynamic power profiles of the electric loads in terms of dynamic power ramp up/down and minimum time on/off of each load is harder to implement in a standard MPC framework. \\
\indent In this paper we define load scheduling as the optimal on/off combinations and timing of a set of distinct electric loads via the computation of an optimal binary control signal $w(k) \in \{0,1\}$. The work is partially motivated by previous work \cite{c10} and \cite{myasme} in which the design and sizing problem of a standalone photovoltaic reverse osmosis (RO) system is considered, where the RO loads are to be scheduled on/off. The work in \cite{c10} computes the optimal size and number of units for a selected location but lacks the procedure for optimally scheduling dynamic loads. Here we aim to find optimal load scheduling by on/off switching of possibly non-linear dynamics electric loads.\\
\indent The MPC optimization problem becomes untractable for binary load switching because of the exponentiation growth of the binary combinations in the length of the prediction horizon $N$ and the number $n$ of loads. Constraints on the allowable load switching help to alleviate the combinatorial problem, making MPC optimization with binary switching computationally feasible.  The approach presented in this paper will be illustrated in a simulation study in which each load has its own dynamics for both turning on and shutting off. Solar forecasting data on a partly cloudy morning and a clear afternoon at UC San Diego is used to illustrate how loads are scheduled to turn on/off dynamically to track solar power predictions. The moving horizon nature of forecasts serve as an ideal input to the MPC algorithm. With the finite prediction horizon in MPC it is crucial to have reliable forecast of power delivery. However, in reality cloud advection forecasts become less accurate over longer horizons as the cloud dynamics render the basic assumptions of static clouds invalid. This inaccuracy leads to error in the MPC decision. Either an overprediction can cause loads to activate during a period where there is not enough solar energy to meet the demand or an underprediction can prevent loads from being schedule even though energy would have been available. For this reason, we will investigate mis-scheduling due to forecast errors.\\
\indent Section~\ref{sec:SDLM} introduces the dynamic loads assumptions in terms of dynamic power ramping and on/off time constraints. Section~\ref{sec:DLS} gives the approach for dynamic load scheduling based on power tracking over a moving prediction horizon of $N$ points, with an admissible set of binary switching combinations. In Section~\ref{sec:AEx} different solar forecast methods are considered and  the effect of forecast errors on the load scheduling is investigated. Advective, persistence and perfect forecasts are used as inputs to the load scheduling algorithm to show how forecasting errors can lead to scheduling errors. The paper is ended by concluding remarks in Section~\ref{sec:CCLS}.
\section{Switched Dynamic Load Modeling}\label{sec:SDLM}
\subsection{Assumptions on Loads}
We consider a fixed number of $n$ loads where the power demand $p_i(t)$, $i=1,2,\ldots,n$ as a function of time $t$ for each load $i$ is modeled by a known switched dynamic system. For the dynamic scheduling of the loads, loads are assumed to be switched ``on'' or ``off'' by a binary switching signal $w_i(t) = \{0,1\}$.

Each load $i$ is also assumed to have a known minimum duration $T_i^{\mbox{\tiny off}}>0$ for the ``off'' time of the load when $w_i(t)=0$ and a minimum duration $T_i^{\mbox{\tiny on}}>0$ for the ``on'' time of the load when $w_i(t)=1$. The duration times $T_i^{\mbox{\tiny off}}$ and $T_i^{\mbox{\tiny on}}$ avoid unrealistic on/off chattering of the switch signal $w_i(t)$ during load scheduling and limit the number of transitions in $w_i(t)$ over a finite optimization period $T>0$. This can be also an equipment safety or operational constrains. 

\subsection{Admissible Switching Signals}

With the minimum on/off duration times $T_i^{\mbox{\tiny on}},T_i^{\mbox{\tiny off}}$ and the finite time period $T$ for load switching, on/off switching of a load at time $t=\tau_i$ can now be formalized. Special care should be given to turning on loads at $t= \tau_i$ close to the final time $\tau_i=T$ which is also depends on the equipments or loads type. For the formalization, the load switching signal $w_i(t)$ will be a combination of an ``on'' signal $w_i^{\mbox{\tiny on}}(t) \in {0,1}$ and an ``off'' signal $w_i^{\mbox{\tiny on}}(t) \in {0,1} $ that both take into account the constraints of minimum on/off duration and the finite time $T$ for load switching. As a result, the admissible on/off transition signal $w_i(t) = \{w_i^{\mbox{\tiny on}}(t),w_i^{\mbox{\tiny off}}(t)\}$ of a load at time $t=\tau_i$ can now be formalized by the switching signal\vspace{-2pt}
\begin{equation}
w_i^{\mbox{\tiny on}}(t) = 
\left \{ 
\begin{array}{rclcl} 
0 &\mbox{for}& t < \tau_i & \mbox{and}& \tau_i \geq T_{i,last}^{\mbox{\tiny off}} + T_i^{\mbox{\tiny off}} \\ 
1 &\mbox{for}&  t \geq \tau_i & \mbox{and} & \tau_i \leq T - T_i^{\mbox{\tiny on}}
\end{array} \right .
\label{eq:wion}\vspace{-3pt}
\end{equation}
where $T_{i,last}^{\mbox{\tiny off}}$ denotes the most recent (last) time stamp at which the load $i$ was switched ``off'',
and the opposite goes to $w_i^{\mbox{\tiny off}}(t)$ \cite{ACC}.



\subsection{Dynamic Load Models}

For the computational results presented in this paper, linear first order continuous-time dynamic models will be used to model the dynamics of the power demand of the loads. It should be pointed out that the computational analysis is not limited to the use of linear first order models, as long as the dynamic models allow the numerical computation of power demand $p_i(t)$ as a function of the switching signal $w_i(t)$.

To allow different dynamics for the time dependent power demands $p_i(t)$ when the binary switching signal $w_i(t) = \{0,1\}$ transitions from 0 to 1 ("on") or transitions from 1 to 0 ("off"), different time constants are used in the first order models. This allows power demands $p_i(t)$ to be modeled at different rates when switching loads. Referring back to the admissible on/off transition signals $w_i^{\mbox{\tiny on}}(t)$ and $w_i^{\mbox{\tiny off}}(t)\}$ respectively in \ref{eq:wion}, the switched linear first order continuous-time dynamic models for a particular load are assumed to be of the form\vspace{-2pt}
\begin{equation}
\alpha_i^{\mbox{\tiny on}} \frac{d}{dt} p_i(t) + p_i(t) = x_i w_i(t),~ p_i(T) = p_i^{\mbox{\tiny off}} ~\mbox{and}~ w_i(t) = w_i^{\mbox{\tiny on}}(t)
\label{eq:odeon}
\end{equation}
to model the power demand $p_i(t)$ of a load. Same goes to the off signal model but different time constants $\alpha_i^{\mbox{\tiny on}}$ and $\alpha_i^{\mbox{\tiny off}}$ are used to model respectively the on/off dynamic switching of the load. 

\section{Discretization and Dynamic Load Scheduling}\label{sec:DLS}
\subsection{Discretization of Models}
To achieve the optimal switching times $\tau_i$ of the binary switching signals $w_i(t)$ for each load discretizing the power demand $p_i(t)$ and the optimal switching signal $w_i(t)$ was performed at a time sampling\vspace{-8pt}
\begin{equation}
\tau_i = N_i \Delta_t
\label{eq:Ni}\vspace{-8pt}
\end{equation}
where $\Delta_t$ is the sampling time and $k=0,1,\ldots$ is an integer index. To simplify the integer math, we assume that both the switching times and the minimum on/off duration times are all multiple of the sampling time $\Delta_t$. With the imposed time discretization, the switching signal $w_i(t_k)$ is held constant between subsequent time samples and $t_k$ and $t_{k+1}$. A Zero Order Hold (ZOH) discrete-time equivalent of the continuous-time models given earlier in \ref{eq:odeon} was used to achieve the computation of $p_i(t_k)$, and it is given by \vspace{-4pt}
\[
p_i(t_k) = b_i^{\mbox{\tiny on}} w_i(t_{k-1}) + a_i^{\mbox{\tiny on}} p_i(t_{k-1}),~ p_i(t_{N_i}) = p_i^{\mbox{\tiny off}} \vspace{-6pt}
\]
\[
~\mbox{and}~ w_i(t_k) = w_i^{\mbox{\tiny on}}(t_k) \vspace{-4pt}
\]
for ''on'' switching of the load. The coefficients $b_i$ and $a_i$ in the first order ZOH discrete-time equivalent models are fully determined by the time constants $\alpha_i^{\mbox{\tiny on}}$, $\alpha_i^{\mbox{\tiny off}}$, static load demand $x_i$ and the chosen sampling time $\Delta_t$. 
\subsection{Power Tracking}
Defining the optimization that allows the computation of optimal discrete-time switching signals $w_i(t_k),~ i=1,2,\ldots,n$ for the power demand $p_i(t_k)$ of $n$ loads. Defining a power tracking error\vspace{-15pt}
\begin{equation}
e(t_k) = P(t_k) -\sum_{i=1}^n p_i(t_k)
\label{eq:powererror}\vspace{-8pt}
\end{equation} 
it is clear that computing optimal $w_i(t_k)$ will involve a criterion function and possible constraints on $e(t_k)$ and $w_i(t_k)$ over a (finite) time horizon $k=1,2,\ldots,N$. Choosing $N$ to be large, e.g. $N = T/\Delta_t$ where $T$ is the complete optimization period, results in two major disadvantages.The first one is that the number of possible combinations of the discretized binary switching signal $w_i(t_k)$ grows exponentially with the number of loads $n$ and the number of time steps $N$. Fortunately, this can be significantly reduced by the requirement of minimum on/off duration times $T_i^{\mbox{\tiny on}},T_i^{\mbox{\tiny off}}$ for the loads.As mentioned before, this avoids unrealistic on/off chattering of the switch signal $w_i(t)$ during load scheduling and significantly reduces the number of binary load combinations.
The second disadvantage of choosing $N$ to be large requires the discrete-time power profile $P(t_k)$ to be available over many time samples to plan for optimal load scheduling, which leads to increasingly suboptimal schedules due to increasing solar forecast errors.

\subsection{Admissible Discrete-Time Switching Combinations}

With the imposed time discretization given in \ref{eq:Ni}, and a finite prediction horizon $N$, the admissible on/off transition signal in \ref{eq:wion} reduces to\vspace{-2pt}
\begin{equation}
w_i^{\mbox{\tiny on}}(t_k) = 
\left \{ 
\begin{array}{rclcl} 
0 &\mbox{for}& k < N_i & \mbox{and}& N_i \geq N_{i,last}^{\mbox{\tiny off}} + N_i^{\mbox{\tiny off}} \\ 
1 &\mbox{for}&  k \geq N_i & \mbox{and} & N_i \leq N - N_i^{\mbox{\tiny on}}
\end{array} \right .
\label{eq:wiondiscrete}\vspace{-2pt}
\end{equation}
where $N_{i,last}^{\mbox{\tiny off}}$ now denotes the most recent discrete-time index at which the load $i$ was switched ``off''. Similarly $w_i^{\mbox{\tiny off}}(t_k) $ reduces to
\begin{equation}
w_i^{\mbox{\tiny off}}(t_k) = 
\left \{ 
\begin{array}{rclcl} 
1 &\mbox{for}& k < N_i & \mbox{and}& N_i \geq N_{i,last}^{\mbox{\tiny on}} + N_i^{\mbox{\tiny on}} \\ 
0 &\mbox{for}&  k \geq N_i
\end{array} \right .
\label{eq:wioffdiscrete}\vspace{-4pt}
\end{equation}
where $N_{i,last}^{\mbox{\tiny on}}$ denotes the most recent discrete-time index at which the load $i$ was switched ``on''. 
Both signals $w_i^{\mbox{\tiny on}}(t_k)$ in \ref{eq:wiondiscrete} and $w_i^{\mbox{\tiny off}}(t_k)$  in \ref{eq:wioffdiscrete} 
form a set ${\cal W}$ of binary values for admissible discrete-time switching signals defined by
\begin{equation}
{\cal W} = \left \{ 
\begin{array}{c}
w_i(t_k) \in \{w_i^{\mbox{\tiny on}}(t_k),w_i^{\mbox{\tiny off}}(t_k)\} ,\\
 i=1,2,\ldots,n,~ k=1,2,\ldots,N\\
\mbox{where}~
\begin{array}{c}
w_i^{\mbox{\tiny on}}(t_k) \in \{0,1\}~\mbox{given in}~\mbox{\ref{eq:wiondiscrete}
}\\
w_i^{\mbox{\tiny off}}(t_k) \in \{0,1\}~\mbox{given in}~\mbox{\ref{eq:wioffdiscrete}
}
\end{array}
\end{array}
\right \}
\label{eq:Wset}\vspace{-2pt}
\end{equation}

It is beneficial to note that the number of binary elements in the set ${\cal W}$ is always much smaller than $(2^n)^{N-1}$ due to required minimum number of on/off samples $N_i^{\mbox{\tiny on}},N_i^{\mbox{\tiny off}}$ for the loads.
This results shows that constraints on the allowable load switching helps to alleviate the combinatorial problem, making an  optimization with binary switching computationally feasible. As an example, consider the case of $n=3$ loads over a power prediction horizon of $N=6$ samples. Without any requirements on minimum number of on/off samples one would have to evaluate $(2^n)^{N-1} = 32768$ possible combinations of the load switching signal $w_i(t_k) \in \{0,1\}$. Starting at a binary combination with all loads off, e.g $w(0)=[0~0~0]$ and requiring the loads to stay on/off for at least 4 samples reduces the number of possible binary combinations to only $2197$. Clearly, the number of combinations reduces even further for a non-zero initial condition, e.g. $w(0)=[1~0~0]$, where the first load that is switched on is required to stay on over the prediction horizon. It is clear from the above illustrations that the number of admissible binary combinations of $n$ loads over a prediction horizon of $N$ points is in general much smaller than $(2^n)^{N-1}$, making the optimization with binary switching computationally feasible for real-time operation.

\subsection{Moving Horizon Formulation}

The dynamic load scheduling optimization problem is formulated as a moving horizon optimization problem by following the power tracking error defined in \ref{eq:powererror} \vspace{-5pt}
\begin{equation}
\begin{array}{c}w_i(t_m) \\ i=1,2,\ldots,n \\ 
m=k,\ldots,k+N-1\end{array} = \mbox{arg} \min_{w_i(t_m) \in {\cal W}}  f(e(t_l)),
\label{eq:MPC} \vspace{-4pt}
\end{equation} \vspace{-5pt}
where   {$ \hspace{50pt} l=k+1,\ldots,k+N$}
\\ \\
with the admissible set ${\cal W}$ defined in \ref{eq:Wset}. 
Adopting the ideas from MPC, the $N \times n$ dimensional optimal switching signal $w_i(t_m)$ over the optimization horizon $m=k,\ldots,k+N-1$ and the loads $i=1,2,\ldots,n$ is selected by the evaluation of the criterion function $f(e(t_l))>0$ as a function of the power tracking error $e(t_l)$ (in the future) at $l=k+1,\ldots,k+N-1$. Once the optimal switching signal $w_i(t_m) \in {\cal W}$, $m=k,\ldots,k+N-1$ is computed, the optimal signal is applied to the loads {\em only\/} at the time instant $t_k$, after which the time index $k$ is incremented and the optimization in \ref{eq:MPC} is recomputed over the moving time horizon.

As the admissible set ${\cal W}$ defined in \ref{eq:Wset} has a finite and countable number of binary combinations for the switching signal, the optimal value for $w_i(t_m) \in {\cal W}$ is computed simply by a finite number of evaluation of the criterion function $f(e(t_l))>0$. Hence, no (gradient) based optimization is used to compute the final value for $w_i(t_k)$. Possible candidate functions $f(e(t_l))>0$ may include a least squares criterion or may include a barrier function 
\begin{equation}
f(e(t_l)) =\sum_{l=k+1}^{k+N} tr\{e(t_l) e(t_l)^T\} - \ln(c(e(t_l)))
\label{eq:optcrit}\end{equation}
to enforce a positive constraint $e(t_l)>0$. Such constraints may be required to guarantee that the load demand is always smaller then the (predicted) power profile $P(t_k)$ in \ref{eq:powererror}. In this paper we use the quadratic function with a barrier function in \ref{eq:optcrit} to perform tracking of predicted solar power curves by dynamic load switching.
\section{Numerical Results}\label{sec:AEx}
 This algorithm can be implemented for any kind of standalone system (wind, solar or even hybrid) with a forecast tool providing input data. Here, we present a standalone solar system powering 3 units of normalized sizes rated at $x_i=$  60\%, 26\% and 12\% of full solar power. Furthermore, every load has different dynamics for on/off switching modeled by the first order time constants $\alpha_i^{\mbox{\tiny on}}$ and $\alpha_i^{\mbox{\tiny off}}$ similar to the model given in (\ref{eq:odeon}). The first order time constants $\alpha_i^{\mbox{\tiny on}}$ and $\alpha_i^{\mbox{\tiny off}}$ are dependent on the size $x_i$ of the load. As illustrated in figure 3 in \cite{ACC}, the proposed scheduling approach schedules the on/off status of the three different loads in order to capture as much solar energy as possible, \textit{i.e.} to decrease the unutilized (lost) energy. The algorithm is tested against a clear sky model predicted day as well as the real PV forecast recorded on September 09, 2014 by UCSD Sky Imager.
\subsection{Perfect forecast - clear sky model}
\begin{figure}[h]\centering
\includegraphics[width=1\columnwidth]{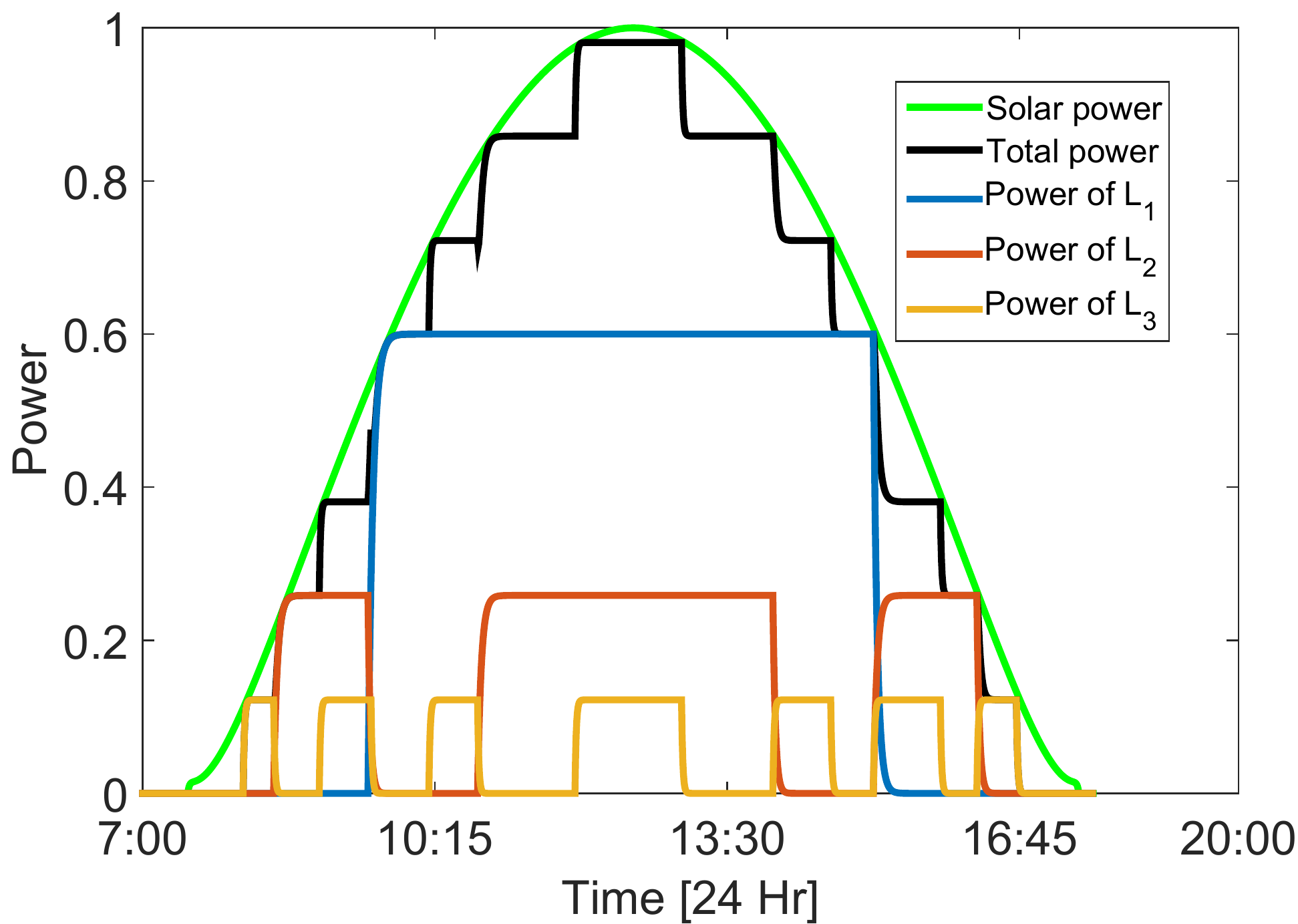}
\caption{Clear day Perfect forecast $N=360$ seconds and $t_k=60$ for three loads.
\href{http://solar.ucsd.edu/c/wp-content/uploads/2015/11/Bellcurvesim1.gif}{Animation can be found here.}}
\label{sim}
\end{figure}

 The solar production from the clear sky model is presented in figure \ref{sim}. As determined in \cite{c10}, when short-term variability is small and deterministic, a smooth scheduling for $n=3$ loads can be achieved, As observed (figure \ref{sim}). 
 It can be seen that the scheduling algorithm emphasizes that turning on the largest unit is the main priority. Lost energy is minimized by combining all three loads through out the day.

Table \ref{forecastscenarios} compares the efficiency of the system, defined as the percentage of solar power the loads capture,   under the different forecast scenarios. A diminishing returns effect is observed for different forecast horizon $N$ and switching time $t_k$. As observed for $N=360$, the efficiency of the system increases as $t_k$ decreases. For $t_k=30$, increases with $N$, but remains constant for $N>30$. From table II we conclude that increasing $N$ is computationally costly and provides diminishing returns for $N$ greater than 8 times the switch time.
\begin{table}
\centering
\caption{Loads characteristics (seconds)}
\label{simu;ationloadchar}
\begin{tabular}{l||l|l|l|l|l|} \hline
\multicolumn{1}{|l||}{\begin{tabular}[c]{@{}l@{}}Loads \\ characteristics\end{tabular}} & Size(\%)& $\alpha_i^{on}$&$\alpha_i^{off}$& $T_i^{on}$&$T_i^{off}$ \\ \hline
\multicolumn{1}{|l||}{$L_1$}& 60& 120&45 & 600& 450\\ \hline
\multicolumn{1}{|l||}{$L_2$}&26&  45& 30& 510& 300\\ \hline
\multicolumn{1}{|l||}{$L_3$}& 12& 15& 15& 450 & 240\\ \hline
\end{tabular}\vspace{-15pt}
\end{table}
\begin{table}[h]
\centering
\caption{Efficiency (\%) results for different forecast horizons. if $t_k$ is not divisible by n, no results are obtained ($\oslash$).}
\label{forecastscenarios}
\begin{tabular}{|l||l|l|l|l|l|}
\hline
$t_k [s]\backslash N [s]$&210&270&360&540&720\\ \hline
30&88.41&89.09&89.09&89.09&89.09\\ \hline 
60 & $\oslash$&$\oslash$ &88.74&88.74&88.74\\ \hline
120& $\oslash$&$\oslash$&87.34&$\oslash$&88.13\\ \hline 
\end{tabular}\vspace{-15pt}
\end{table}

\subsection{Forecast Scenarios - actual data}

Solar forecasting is critical for load scheduling in standalone PV systems due variability of solar resources. UC San Diego has developed a ground based sky imager to detect clouds, cloud velocity, and forecast the advection of cloud shadows on the ground over the coming 10 to 20 minutes \cite{c7}. The imager is composed of an upward facing camera coupled with a fisheye lens to capture a large area of O(10 km2). The forecasting algorithm uses projected cloud locations coupled with a clear sky index model to predict global horizontal irradiance (GHI) over the captured domain. This is referred to as ``advection forecast''.

A commonly implemented forecasting method is to assume that the clear sky index \textit{kt} at time \textit{t} remains constant for the short period of the forecast \textit{t+N}, with irradiance increasing at the clear sky value multiplied by \textit{kt}. This method is called "persistence" forecasting. Any more complex forecasting technique can be benchmarked against these persistence forecasts. Persistence and advection forecast methods are associated with erroneous predictions. Persistence errors are due to the deviation from the initial cloud state, which generally increases in time. Advection errors arise mainly from inaccuracy of cloud detection and mis-prediction of cloud formation, evaporation, or deformation. For this reason, the accuracy of the UCSD solar forecasting algorithm against perfect, and persistence conditions are compared, and their effects on scheduling algorithms is classified. The aggregate forecast errors for the forecast for this day are: $rRMSE=14.5\%$, $rMBE=2.7\%$, $rMAE=5.3\%$, $rMAE_p=7.9\%$ computed by \cite{c7}. 

Figure \ref{Compare} illustrates the results of the MPC scheduling algorithm under the three different forecast scenarios. By design the scheduling under the perfect forecast is error free. However, under persistence and advection forecast predictions we can see several areas where the load demand is greater than the available solar energy.\\
\indent An overprediction can cause loads to activate during a period where there is not enough solar energy to meet the demand (Power exceedence or PE); an underprediction can prevent loads from being scheduled even though energy would have been available resulting in lost energy. Two metrics are considered to evaluate the accuracy of the control algorithms. The first metric is the number of time steps with PE while the second one is called energy exceedence (EE) and is equal to integration of PE at the consecutive violated time steps. The EE provides a reasonable estimate for the size of an energy storage system that would avoid PE.\\
\indent Figure \ref{Compare} depicts that three PE situations happen for the persistence forecast scenario between 9:00 and 11:00. At 9:10, the intermediate sized unit (unit 2) is in PE for approximately 10 minutes. At 10:20, the smallest unit (unit3) is in a shallow PE in two distinct periods over 10 minutes. Finally at 10:45, we see the largest unit is in PE for 5 minutes. Referring to the advective forecast we the same exceedance scenarios at 10:20 and 10:45, but the PE at 9:10 is avoided.\\

\begin{figure}[ht]\centering
\includegraphics[width=\columnwidth]{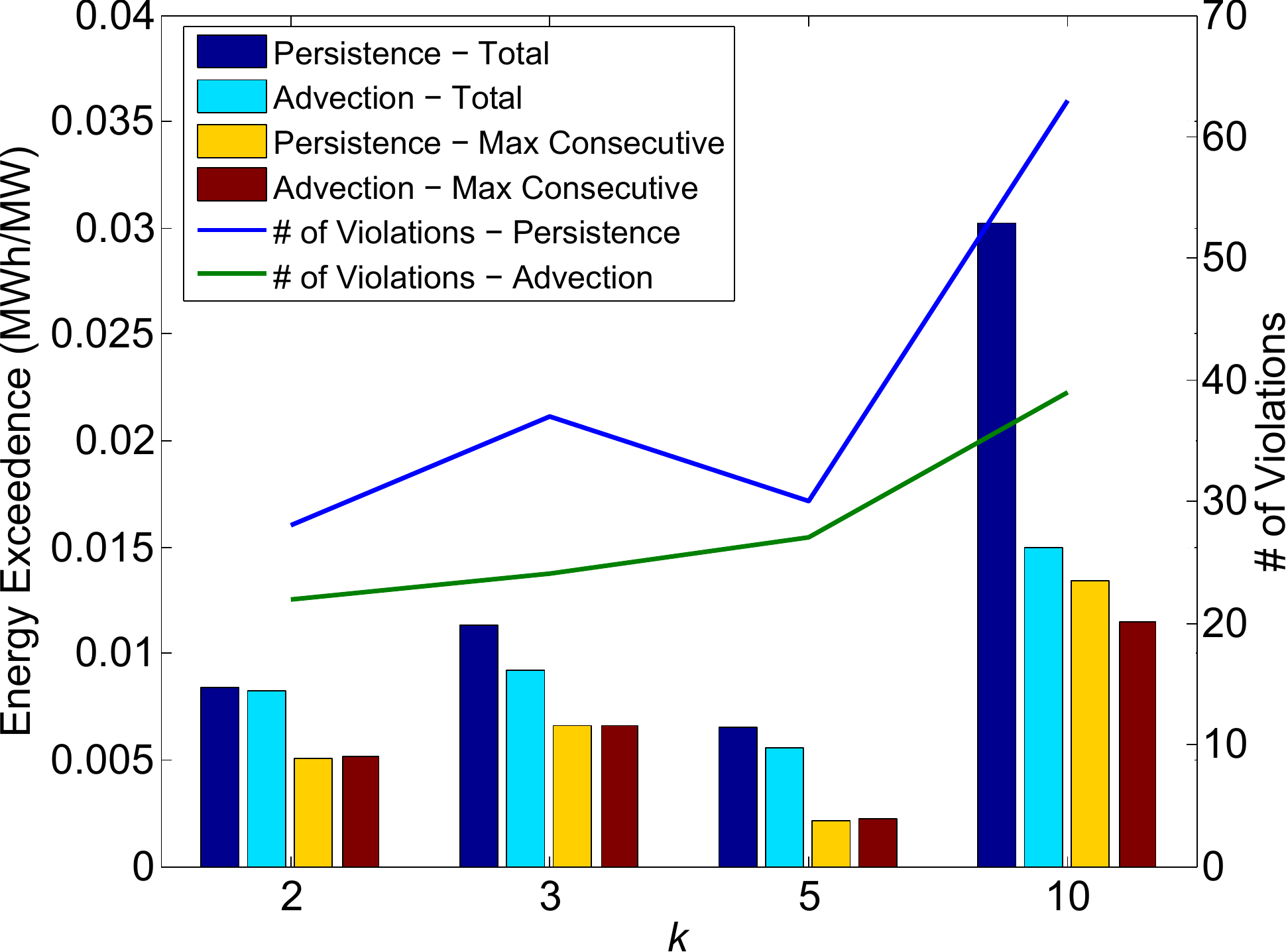}
\caption{Energy exceedance and number of violations for different \textit{k}.}
\label{OverReach}
\end{figure}
\indent Figure \ref{OverReach} shows the total EE for the day,  the maximum energy of individual consecutive PE events, and the total number of violations for the day as a function of \textit{k}. We see that the persistence forecast leads to higher exceedance and more violations for all \textit{k}. Increasing \textit{k} in general also leads to an increase in exceedance and violations. 
\begin{figure}[ht]\centering
\includegraphics[width=\columnwidth]{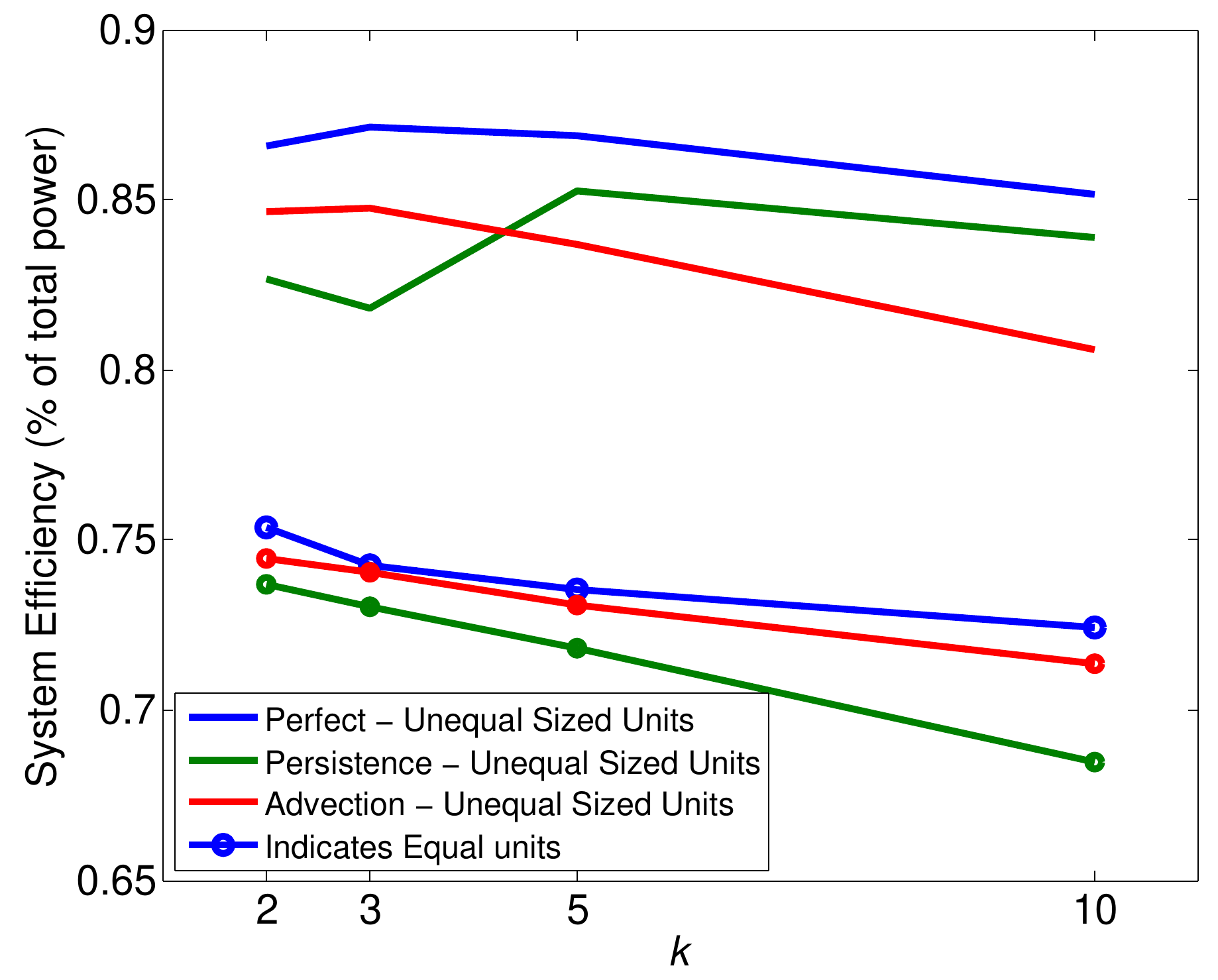}
\caption{Fraction of total solar energy utilized (efficiency) under each forecast scenario for different \textit{k}}
\label{efficiency}
\end{figure}
\indent If we evaluated solar forecast or control algorithms solely based on PE, then a forecast that is biased small (or even always zero) would improve the results as less and/or smaller loads would be scheduled. Therefore a different metric such as efficiency is important to observe. Figure \ref{efficiency} shows the efficiency under each scenario as a function of \textit{k}. It is observed that the perfect forecast has the highest efficiency. For low \textit{k}, the advective forecast captures more energy, where the opposite is true for high k.\\
\begin{figure*}[ht]\centering
\includegraphics[width=4\columnwidth, height=.655\columnwidth,keepaspectratio]{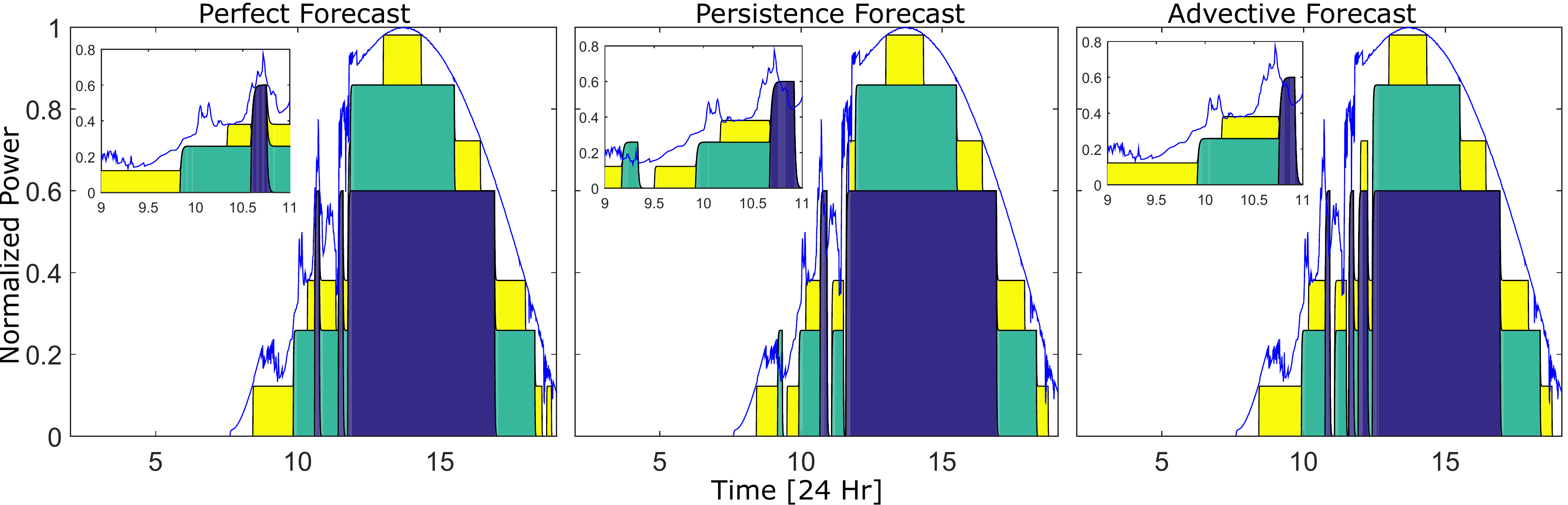}
\caption{Load scheduling for different solar forecast. The scenarios are Perfect forecast (left), persistence forecast (center), and advective forecast (right). The blue area represents $L_3$, green $L_2$ and yellow $L_1$. Each figure includes a subplot showing the region from 9:00 to 11:00, to highlight regions where scheduling errors exist. Power exceedence (PE) errors occur when the colored areas exceed the actually available solar power (shown in blue) as a result of forecast error.}
\label{Compare}\vspace{-15pt}
\end{figure*}
\indent In both figures \ref{OverReach} and \ref{efficiency}, there is non-monotonic behavior associated with increasing \textit{k}, which is not expected. For this reason, the same analysis was run with three units of equal size (\textit{$x_i$}). This is plotted in figure \ref{efficiency}. For this case, the efficiency decreases monotonically with \textit{k} (and is much lower than the non-equal case), and the advective forecast always captures more energy than the persistence forecast. We can conclude that the nonlinear behavior seen in non-equal units across increasing \textit{k} is due to the dynamics associated with starting and shutting down the loads.\\
\indent Energy storage systems such as batteries could help overcome EE episodes as well as increase the efficiency. The excess of energy that is not captured by the loads (figures \ref{Compare},\ref{efficiency}) could be stored and used to power loads during EE events. For battery sizing one could consider match the energy capacity of the battery to the energy required for maximum exceedance of the largest unit over its minimum required on-time. However, figure \ref{OverReach} demonstrates that the maximum individual EE for all cases is a small percentage the power of even the smallest unit. Reducing the size of the battery reduces the up front capital cost associated with the system. A more sophisticated method for battery sizing would be to probabilistically determine the maximum individual EE case for several years of historical forecast data and size accordingly, which will be the focus of future work.

\section{Conclusions}\label{sec:CCLS}
In this paper, a MPC model was developed to compute the optimal binary control signal by determining the on/off combinations and timing of a set of distinct electric dynamic loads scheduling. The MPC load scheduling algorithm was tested using different forecasting techniques to assess the effects of input inaccuracy. The algorithm worked optimally under a perfect forecast (as designed), but created errors due to forecasting error. The energy captured by the loads decreased for increasing \textit{k} in a non-monotonic manner for optimally sized unequal units. However, for the same scenario with equally (but not optimal) sized units, efficiency decreased monotonically with increasing \textit{k} for all forecast scenarios. The advection forecast model created fewer errors and gave less total exceedance and number of violations independent of the \textit{k}, as compared to persistence. For a stand alone system these errors could be mitigated with storage capacity to provide the power to the load to ride through the exceedance period. A battery sizing method is discussed as the topic of future work.

\end{document}